\documentclass[prb,twocolumn,aps,superscriptaddress]{revtex4-2}
\usepackage{graphicx}
\usepackage{dcolumn}
\usepackage{bm}
\usepackage{amssymb}
\usepackage{amsmath}
\usepackage{subfigure}
\usepackage{physics}
\usepackage{sublabel}
\usepackage[utf8]{inputenc}
\usepackage{hyperref}
\usepackage[usenames, dvipsnames]{color}
\usepackage[english]{babel}
\usepackage[T1]{fontenc}
\usepackage{bbm}

\usepackage{float}
\hypersetup{colorlinks = true, linkcolor=blue, citecolor=blue, urlcolor=blue}

\begin{document}

\title{Robust all-electrical topological valley filtering using monolayer 2D-Xenes}

\author{Koustav Jana}
%  \email{koustavjana@ee.iitb.ac.in}
\author{Bhaskaran Muralidharan}%
%  \email{bm@ee.iitb.ac.in}
\affiliation{Department of Electrical Engineering, Indian Institute of Technology Bombay, Powai, Mumbai-400076, India
}%

\date{\today}
% \begin{tocentry}

% Some journals require a graphical entry for the Table of Contents.
% This should be laid out ``print ready'' so that the sizing of the

% Inside the \texttt{tocentry} environment, the font used is Helvetica
% 8\,pt, as required by \emph{Journal of the American Chemical

% The surrounding frame is 9\,cm by 3.5\,cm, which is the maximum
% permitted for  \emph{Journal of the American Chemical Society}
% graphical table of content entries. The box will not resize if the
% content is too big: instead it will overflow the edge of the box.

% This box and the associated title will always be printed on a
% separate page at the end of the document.
% \begin{figure}
%     \centering
%     \includegraphics{graphical_abstract.png}
%     \caption{Caption}
%     \label{fig:my_label}
% \end{figure}
% \includegraphics[width=\textwidth]{graphical_abstract.pdf}
% \end{tocentry}

\begin{abstract}
%   This is an example document for the \textsf{achemso} document
%   class, intended for submissions to the American Chemical Society
%   for publication. The class is based on the standard \LaTeXe\
%   \textsf{report} file, and does not seek to reproduce the appearance
%   of a published paper.

%   This is an abstract for the \textsf{achemso} document class
%   demonstration document.  An abstract is only allowed for certain
%   manuscript types.  The selection of \texttt{journal} and
%   \texttt{manuscript} will determine if an abstract is valid.  If
%   not, the class will issue an appropriate error.

We propose a realizable device design for an all-electrical robust valley filter that utilizes spin protected topological interface states hosted on monolayer group-IV 2D-Xene materials with large intrinsic spin-orbit coupling.
In contrast with conventional quantum spin-Hall edge states localized around the $X$-points, the interface states appearing at the domain wall between topologically distinct phases are either from the $K$ or $K’$ points, making them suitable prospects for serving as valley-polarized channels.
We show that the presence of a large band-gap quantum spin Hall effect enables the spatial separation of the spin-valley locked helical interface states with the valley states being protected by spin conservation, leading to a robustness against short-range non-magnetic disorder.
By adopting the scattering matrix formalism on a suitably designed device structure, valley-resolved transport in the presence of non-magnetic short-range disorder for different 2D-Xene materials is also analyzed in detail.
Our numerical simulations confirm the role of spin-orbit coupling in achieving an improved valley filter performance with a perfect quantum of conductance attributed to the topologically protected interface states. Our analysis further elaborates clearly the right choice of material, device geometry and other factors that need to be considered while designing an optimized valleytronic filter device.

\end{abstract}

\maketitle
\section{Introduction}
% This is a paragraph of text to fill the introduction of the
% demonstration file.  The demonstration file attempts to show the
% modifications of the standard \LaTeX\ macros that are implemented by
% the \textsf{achemso} class.  These are mainly concerned with content,
% as opposed to appearance.

\begin{figure*}
    \centering
    \includegraphics[scale=0.55]{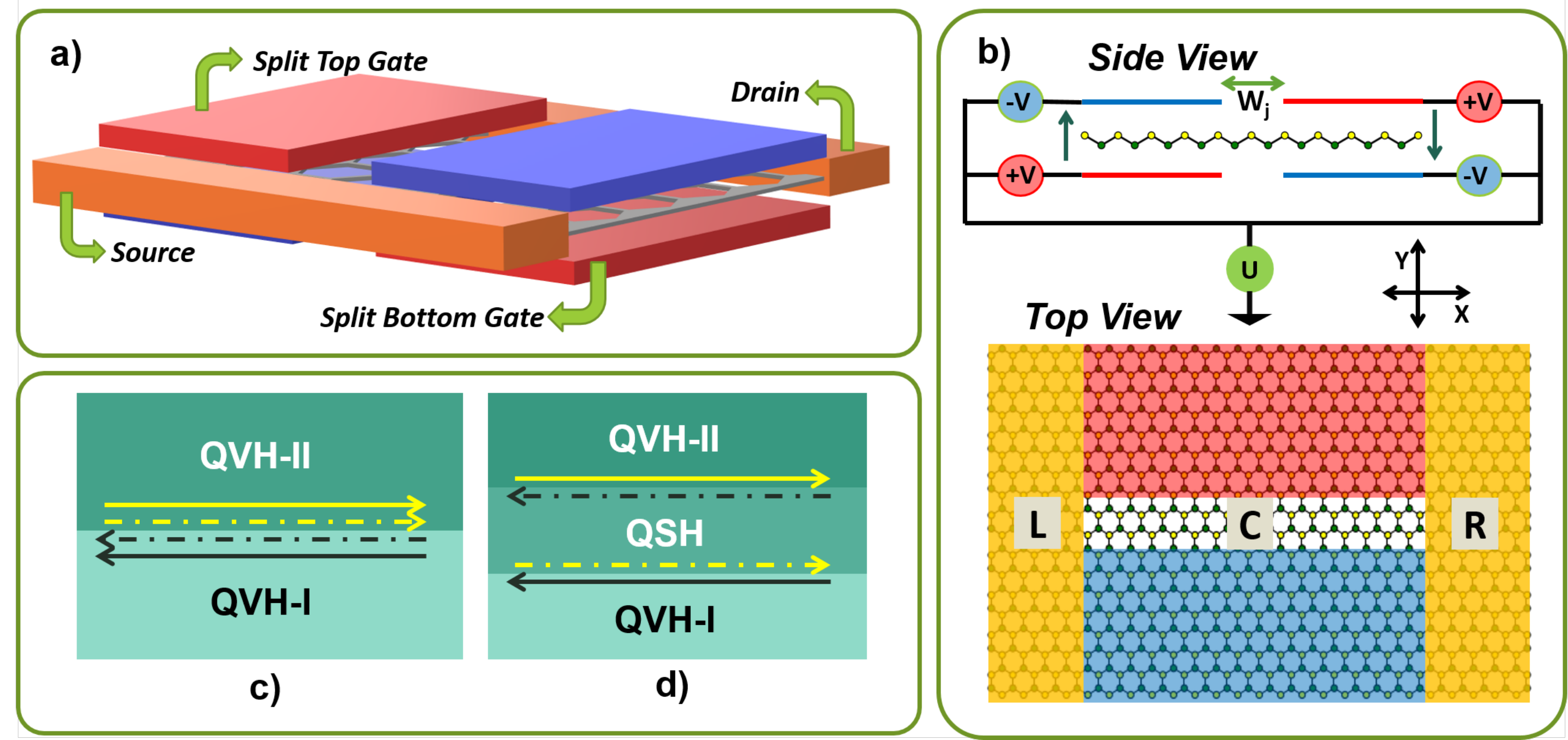}
    \caption{Device schematics:
    \textbf{a)} Proposed dual-split-gated device where red denotes the positive plate and blue denotes the negative plate. A monolayer 2D Xene nanoribbon with a buckling height $2l$ is sandwiched between the gates. This configuration helps us achieve opposite electrical field $E_{z}$ on either sides with a line junction in between where $E_{z}$ transitions from positive to negative or vice-versa.
    \textbf{b)} Side and top view of the device structure. Sublattice $A$ sites are marked in yellow and sublattice $B$ sites in green. The width of the line junction is determined by the spacing $W_{j}$ between the adjacent gates. The location of the Fermi level of the line junction can be controlled by $U$ whereas $E_{z}$ can be controlled by V. The left lead, the right lead and the channel region are denoted by $L$,$R$ and $C$ respectively.   
    \textbf{c)} Earlier proposals feature domain walls between the QVH phases with reversed inversion asymmetry, giving rise to valley-momentum locked interface states for both the spins.
    \textbf{d)} Our proposal features domain walls between the QSH and the QVH phases spatially separated from each other, such that, each one hosts spin-valley locked interface states and hence the valley-polarized conducting electrons are spin-protected when TRS is preserved. The yellow (dark blue) line denotes $K$ ($K'$) valley whereas the solid (dotted) line denotes the up(down) spin states. }
    \label{fig:dual_gate}
\end{figure*}

\label{section:Introduction}
% \texttt{Cover aspects of importance of valley filtering. Proposals on bilayer and their robustness. Introducing topological robustness and how our proposal outweights the other issues.}
% Topological interface states at the line junctions in monolayer and bilayer graphene, although have shown promise in realising a perfect valley filter, suffer from deteriorated efficiency in presence of strong short-range disorder. 
Two-dimensional (2D) materials beyond graphene~\cite{graphene} featuring honeycomb lattices, such as MoS\textsubscript{2} and other transition metal di-chalcogenides (TMDCs) ~\cite{mos2}, group-IV 2D-Xenes such as silicene, germanene, stanene~\cite{silicene,si_ge_sn} etc.,  have enjoyed significant research activity targeting a wide range of applications.
The uniqueness of these materials lies in their bandstructure having energy minimas far apart in the momentum space, endowing the low-energy carriers a valley degree of freedom that can be exploited for information manipulation. This has paved the way for the field of valleytronics~\cite{valley_review1,valley1,valley_opto,Cao2012_valleymos2,Sui2015_BLGvalley,Zeng2012_valleymos2,Mak2012_valleymos2,beenakker2007,vhe1_mak2014,vhe2_Shimazaki2015}, in which the valley filter~\cite{beenakker2007} is a primary device paradigm that facilitates the generation of spatially separated valley-polarized carriers. \\
\indent There have been several proposals for valley filtering that include utilizing nanoconstrictions~\cite{beenakker2007}, optical pumping~\cite{Zeng2012_valleymos2,Cao2012_valleymos2,Mak2012_valleymos2,vhe1_mak2014,Gopal}, the valley Hall effect~\cite{vhe1_mak2014,valley1,valley_opto,vhe2_Shimazaki2015,Sui2015_BLGvalley}, strain engineering~\cite{strain2_colin,strain1_jiang2013}, the valley polarized quantum anomalous Hall (QAH) phase~\cite{honeycomb_modes,valleyQAH, valleyQAH_zhang2011}, and domain walls between materials with broken inversion symmetry~\cite{topo2d_review1,Ezawa_2012,dacosta2015,junzhu2016,cheng2016,qah_qvh_valley,top_dom_wall,Liu2016,SOC_DW_Abergel_2014,yang2020,zhang2020,wang2014,MLGdomain_Semenoff2008,MLGmultidomain_Liu2013}. Two indices critical for a valley filter performance are the valley polarization and the total transmission, and their immunity against backscattering~\cite{honeycomb_modes}.
\\
\indent Interface states at domain walls in monolayer 2D materials created using line junctions or defects which have been considered in previous works have different levels of topological robustness ~\cite{topo2d_review1,dacosta2015,junzhu2016,cheng2016,qah_qvh_valley,top_dom_wall,Liu2016,SOC_DW_Abergel_2014,yang2020,zhang2020,wang2014,MLGdomain_Semenoff2008,Jan_1,Jan_2,MLGmultidomain_Liu2013}. Interface states formed along the zero mass lines, where the effective mass reverses sign, are valley-momentum locked and hence serve as perfectly valley polarized channels.    
In buckled honeycomb lattices~\cite{Ni2012_SiGe} of 2D-Xenes and bilayer materials such as bilayer graphene (BLG) ~\cite{dacosta2015,junzhu2016,cheng2016,Sui2015_BLGvalley} and bilayer MoS\textsubscript{2}~\cite{bilayermos2_Wu2013}, it is possible to break the inversion symmetry and control the band gap by the application of a perpendicular electric field using electrical gating, in order to possibly facilitate an all-electrical valley filter. Previous proposals and experimental realizations of domain wall-based valley filters in monolayer \cite{MLGdomain_Semenoff2008,MLGmultidomain_Liu2013} and bi-layer structures ~\cite{junzhu2016,dacosta2015} suffer from a serious deterioration in the transmission and valley polarization due to back-scattering and bulk-assisted inter-valley scattering from strong short-range disorder~\cite{cheng2016,junzhu2016,wang2014}.\\
\indent In this paper, we utilize the spin protection to the valley states through spin-valley locking \cite{top_dom_wall}, to demonstrate a realizable device design for robust valley filtering via the spatial separation of spin-valley locked interface states. This is, in general possible in materials with strong spin-orbit (SO) coupling featuring a broken inversion symmetry, typical examples being monolayer MoS\textsubscript{2} and related materials~\cite{spinvalley_mos2}, as well as group-IV 2D-Xenes under a  perpendicular electric field~\cite{ezawa2012,ezawa2015}. Despite naturally possessing broken inversion symmetry, the band gap in monolayer TMDCs cannot be controlled via electrical gating~\cite{bilayermos2_Wu2013}. Hence, we center the proposal on monolayer group-IV 2D-Xenes with buckled lattice structures \cite{Sireview1,Sireview2,Sireview3,Ni2012_SiGe,Tao2015_SiFET,Ge_Bampoulis_2014,Ge_Derivaz2015,Sn_Zhu2015} to design and realize topologically robust valley filtering. \\
\indent Our device design is inspired by the following ideas. Buckled 2D-Xene materials, in the absence of an electric field possess a topologically non-trivial bandstructure with quantum spin Hall (QSH) edge states and continue to do so until a critical field beyond which they transition to a topologically trivial band insulator phase also known as quantum valley Hall (QVH) phase~\cite{ezawa2015}. Adopting the experimentally well-established dual-split gate structure~\cite{junzhu2016,dacosta2015} as depicted in Fig.~\ref{fig:dual_gate}(a), with the side and top views as shown in Fig.~\ref{fig:dual_gate}(b), we demonstrate that it is also possible to achieve QVH-QSH-QVH domain walls in contrast with the QVH-QVH domain walls of the earlier proposals.
The QVH-QVH domain wall as illustrated in Fig.~\ref{fig:dual_gate}(c) hosts valley-momentum locked states, for both the spins, operating as valley-polarized channels.
On the other hand, with the QVH-QSH-QVH domain walls illustrated in Fig.~\ref{fig:dual_gate}(d), we can obtain spatially separated valley-polarized channels protected by spin conservation through spin-valley locking, provided time reversal symmetry (TRS) is preserved. This introduces the desired spin protection to the valley states via spin-valley locking in each of the two domain walls. Given the two domain walls are spatially well separated, we expect our valley filter to be robust, even against short-range disorder that can cause large momentum transfer, given the disorder is non-magnetic and does not break TRS. This demands a QSH region that has a large gap, i.e., a large intrinsic SO coupling. \\

    % \textbf{a)} Side view of the dual-split-gate device proposed with red denoting the positive plate and blue denoting the negative plate. A monolayer 2D Xene nanoribbon with a buckling height $2l$ is sandwiched between the gates. Sublattice $A$ sites are marked in yellow and sublattice $B$ sites in green. This configuration helps us achieve opposite electrical field $E_{z}$ on either sides with a line junction in between where $E_{z}$ transitions from positive to negative or vice-versa. The width of the line junction is determined by the spacing $W_{j}$ between the adjacent gates. The location of the Fermi level of the line junction can be controlled by $U$. 
    % \textbf{b)} Top view of the device. The left lead, the right lead and the channel region are denoted by $L$,$R$ and $C$ respectively.

\indent We also demonstrate this high degree of robustness by subjecting our channel to short-range non-magnetic disorder of varying strength and examining the effect of SO coupling and line junction width on the valley filter performance. Based on our findings we conclude the superiority of our QVH-QSH-QVH structure compared to other existing proposals. This implies the preservation of unity transmission and a mild degradation of the valley polarization with increase in disorder, guaranteeing a large enough valley polarized current. To support our claims, we also present the local density of states (LDOS) calculations over the entire channel region. Apart from demonstrating the role of SO coupling and line junction width in achieving improved valley filtering, we also present a strategy to optimize the performance of the valley filter to tap into the full potential of our design.
%\indent We organize the rest of the paper in the following way. The set up along with the related physics of topological domain walls and their role in valley filtering is detailed in Sec.~\ref{section:Setup}. The numerical methods as well as a description of the disorder added is presented in Sec.~\ref{section:Methodology}. In Sec.~\ref{section:Results}, we present the main results along. with the optimization strategy for best valley filter performance.  We end Sec.~\ref{section:Results} with a discussion on the experimental feasibility of our design. In Sec.~\ref{section:Conclusion} a brief conclusion is given.

\section{Results and Discussions}
\subsection{Device set up}
\label{subsection:Setup}
In our proposed valley filter device sketched in Figs.~\ref{fig:dual_gate}(a) and~\ref{fig:dual_gate}(b), we consider the channel as well as the leads to be made of monolayer 2D-Xene. The tight binding Hamiltonian, based on the Kane-Mele model~\cite{kane_mele}, for a typical monolayer 2D-Xene having a honeycomb lattice structure reads ~\cite{ezawa2015,ezawa2012}
% \begin{equation} \label{Heq}
% \hat{H}  = -t\sum_{\langle i,j \rangle\alpha}^{} c_{i\alpha}^{\dagger}c_{j\alpha} +
%             i \frac{\lambda_{SO}}{3\sqrt{3}}\sum_{\langle \langle i,j \rangle \rangle \alpha}^{} \alpha \nu_{ij}c_{i\alpha}^{\dagger}c_{j\alpha} 
%             +\sum_{i\alpha}^{}\mu_{i}\Delta_{z}c_{i\alpha}^{\dagger}c_{i\alpha}
% \end{equation}
\begin{multline} \label{Heq}
\hat{H}  = -t\sum_{\langle i,j \rangle\alpha}^{} c_{i\alpha}^{\dagger}c_{j\alpha} +
            i \frac{\lambda_{SO}}{3\sqrt{3}}\sum_{\langle \langle i,j \rangle \rangle \alpha}^{} \alpha \nu_{ij}c_{i\alpha}^{\dagger}c_{j\alpha} \\
            +\sum_{i\alpha}^{}\mu_{i}\Delta_{z}c_{i\alpha}^{\dagger}c_{i\alpha},
\end{multline}
where $c_{i\alpha}^{(\dagger)}$ represents the annihilation (creation) operator of an electron on site $i$ with a spin $\alpha$, and
$\langle i,j \rangle$ and $\langle \langle i,j \rangle \rangle$ run over all the nearest and next-nearest neighbour hopping sites respectively. The spin index $\alpha$ can be $\uparrow/ \downarrow$, represented with corresponding values $+1/-1$ respectively.
The first term in \eqref{Heq} corresponds to the usual nearest neighbour hopping term with a hopping strength $t$. 
The second term represents the intrinsic SO coupling with strength $\lambda_{SO}$, where $\nu_{ij} = +1(-1)$ for anti-clockwise(clockwise) next-nearest neighbour hopping with respect to the positive $z$-axis.
The third term denotes the staggered sublattice potential due to $\Delta_{z}$, with $\mu_{i} = +1(-1)$, when $i$ belongs to sublattice A(B). Here, $\Delta_{z}$ is site dependent even though the $i$ index has been dropped. This term can be easily realized in buckled lattices through the application of a perpendicular electric field $E_{z}$, giving $\Delta_{z} = lE_{z}$ where $2l$ is the buckling height.There are additional Rashba SO coupling terms which have been neglected because they either are negligible in value for group-IV Xenes or have negligible effect on the states around the $K$/$K'$ points~\cite{ezawa2015,ezawa2012,zhang2020,dimi2021,Ezawa_2012}.\\
\begin{table}[b]
\caption{The parameters corresponding to graphene, silicene, germanene and stanene. $t$ is the hopping parameer, $a$ is the lattice constant, $\lambda_{SO}$ is the intrinsic SO coupling, $l$ is half the buckling height~\cite{ezawa2015}}
    \centering
    \begin{tabular}{c c c c c c c c c}
    \hline
     & & $t(eV)$ & & $a$({\AA}) & & $\lambda_{SO}(meV)$ & & $l$({\AA})  \\
    \hline     
    Graphene & & 2.8 & & 2.46 & & $10^{-3}$ & & 0  \\
    Silicene & &1.6 & & 3.86 & & 3.9 & & 0.23 \\
    Germanene & &1.3 & & 4.02 & & 43 & & 0.33 \\
    Stanene & & 1.3 & & 4.7 & & 100 & & 0.4 \\
    \hline
    \end{tabular}
    \label{tab:Xene}
\end{table}
\begin{table}[b]
\caption{Chern numbers ($\mathcal{C}^{\eta}_{s}$) corresponding to the spin ($s$) and valley ($\eta$) degrees of freedom, along with the spin ($\mathcal{C}_{s}$) and valley ($\mathcal{C}_{v}$) Chern numbers, for the QSH, QVH-I and QVH-II phases~\cite{ezawa2015,ezawa2012,top_dom_wall,zhang2020, wang2014,Liu2016}}
    \centering
    \begin{tabular}{c c c c c c c c c c c c c}
    \hline
     & & $\mathcal{C}^{K}_{\uparrow}$ & & $\mathcal{C}^{K}_{\downarrow}$ & & $\mathcal{C}^{K'}_{\uparrow}$ & & $\mathcal{C}^{K'}_{\downarrow}$ & & $2\mathcal{C}_{s}$ & & $\mathcal{C}_{v}$ \\
    \hline     
    QVH-I & & 1/2 & & 1/2 & & -1/2 & & -1/2 & & 0 & & 2\\
    QSH & & 1/2 & & -1/2 & & 1/2 & & -1/2 & & 2 & & 0  \\
    QVH-II & & -1/2 & & -1/2 & & 1/2 & & 1/2 & & 0 & & -2\\
    \hline
    \end{tabular}
    \label{tab:Chern}
\end{table}
\textbf{Choice of the material.}
Typical values of the parameters involved in (\ref{Heq}) for the candidate materials~\cite{ezawa2015} have been summarized in Tab.~\ref{tab:Xene}.
Graphene does not have a buckled structure ($l=0$) and hence electrical gating cannot be an option to introduce the $\Delta_{z}$ term in it. Also both graphene and silicene have very small $\lambda_{SO}$ values of $10^{-3}meV$ and $3.9meV$ respectively, thus not making them suitable candidates to exploit the topological robustness of the QSH phase that we intend to do in order to design an efficient valley filter. It is worth mentioning that theoretical possibilities of engineering QSH and QVH states in strained graphene exist \cite{strain2_colin}, even in the absence of SO coupling, but such proposals rely on complex quantum pumping processes. \\
\indent Stanene with an intrinsic SO coupling of $\approx 0.1eV$, seems to be a promising candidate for our valley filter~\cite{ezawa2015, SiGeSn_SOC} proposal. Apart from this, functionalized germanene~\cite{Ge_large_gap} and stanene~\cite{Sn_large_gap,Sn_large_gap_ethnyl} have shown sizable band gaps of around $0.3eV$ along with a record band gap of around $1.34eV$ in chemically decorated monolayer plumbene~\cite{plumbene_zhao2016}. Also the above materials show a large enough buckling height~\cite{plumbene_zhao2016,Sn_large_gap_ethnyl}, with $l$ ranging from $0.4-0.7${\AA} and hence the possibility of achieving a large gap QVH state as well. Thus the possibility of large band gap QSH and QVH phases in germanene and stanene makes them the ideal choice of materials to serve our purpose.\\ 
% On the other hand, germanene and stanene have large band gap QSH phases, owing to their large enough $\lambda_{SO}$ values, and an ample buckling height, making them the perfect choice of materials to serve our purpose. \\
\indent Having $l=0.4$ {\AA} would require an electrical field ($E_{z}$) of $5V/nm$ to achieve $\Delta_{z} = 0.2eV$, which is practically quite achievable~\cite{sakanashi_Efield,Ni2012_SiGe}.
Based on this discussion, we choose the parameters involved in our model to be the following: $t=1.3eV$, $l = 0.4${\AA} and $a = 4${\AA}.
We then vary $\lambda_{SO}$ and $\Delta_{z}$ within the practically realizable limits, in order to optimize the valley filter performance.\\
\begin{figure*}
    \centering
    \includegraphics[scale=0.55]{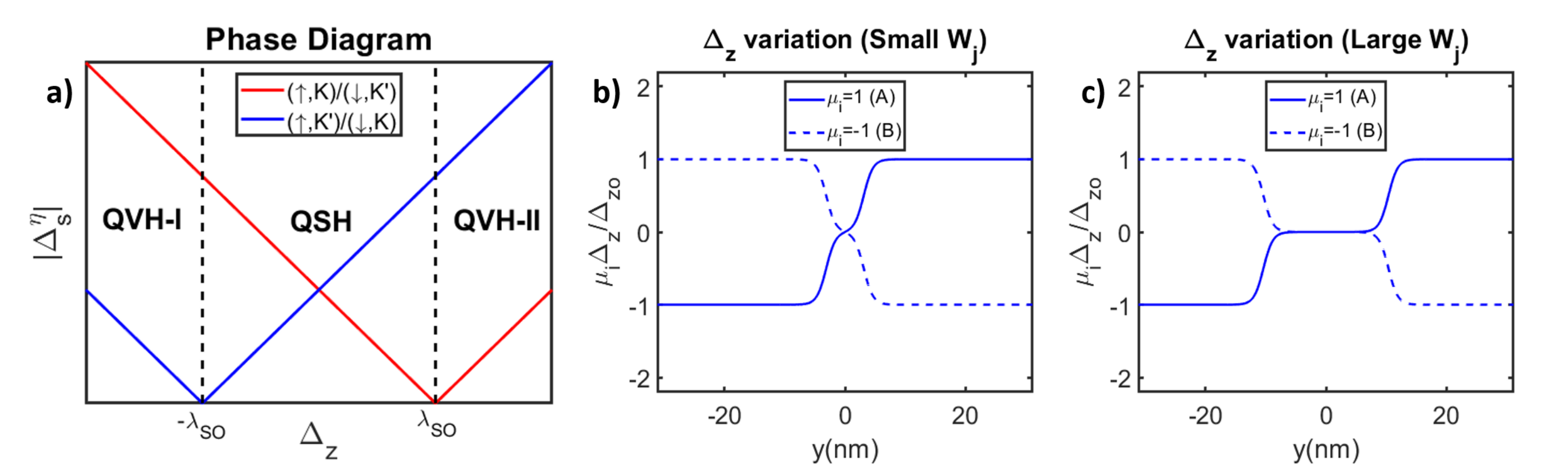}
    \caption{\textbf{a)} Phase diagram depicting the effect of the sublattice potential, $\Delta_{z}$ on the Dirac mass, $|\Delta_{s}^{\eta}|$ (see text). Topological phase transitions at $\Delta_{z} = \pm \lambda_{SO}$. At $\Delta_{z} = - \lambda_{SO}$, the gap closes for $(\uparrow,K’)$ and $(\downarrow,K)$ and we have a QVH-I/QSH transition. Similarly at $\Delta_{z} = \lambda_{SO}$, gap closes for $(\uparrow,K)$ and $(\downarrow,K')$ and we have a QVH-II/QSH transition. \textbf{b)} The staggered sublattice potential, $\mu_{i}\Delta_{z}$ variation along the y-direction in the sublattices(solid line for sublattice A and dashed line for sublattice B) when $W_{j} = 6.25nm$, which is small enough such that the width of the region having $\Delta_{z}=0$ is negligible. $\Delta_{zo}$ is the maximum value of $\Delta_{z}$, which is varied as per our band gap requirements.  
    \textbf{c)} $\mu_{i}\Delta_{z}$  variation when $W_{j} = 20.8nm$, which ensures that the width of the region having $\Delta_{z}=0$ is large enough. } 
    \label{fig:phase}
\end{figure*}
\textbf{Topological Phases.}
Based on the low energy Dirac Hamiltonian \cite{ezawa2012}, the gap of the energy spectrum corresponding to a spin $s$ and valley $\eta$, is given by $2|\Delta^{\eta}_{s}|$, where $\Delta^{\eta}_{s} = \Delta_{z}-\eta s \lambda_{SO}$ is the Dirac mass term~\cite{ezawa2015,ezawa2012}. We define the point $(4\pi/3a,0)$ as the $K$ valley and $(2\pi/3a,0)$ as the $K’$ valley. 
On varying $\Delta_{z}$,  $\Delta^{\eta}_{s}$ reverses its sign when $\Delta_{z}=\eta s \lambda_{SO}$ signalling a topological phase transition.\\
\indent The above phase transitions can be better understood by an analysis of the Berry curvature and the Chern number. 
It is well known that the Berry curvature of the given Hamiltonian \eqref{Heq} is strictly localized around the $K$ and $K’$ points, allowing us to associate each valley with a Chern number $\mathcal{C}^{\eta}_{s}$, where spin index $s=\uparrow,\downarrow$ and valley index $\eta=\pm$.
The Chern number $\mathcal{C}^{\eta}_{s}$ can be obtained using the Dirac Hamiltonian and is given by $\mathcal{C}^{\eta}_{s} = -\frac{\eta}{2}sgn(\Delta^{\eta}_{s})$~\cite{ezawa2015,ezawa2012,top_dom_wall,zhang2020,wang2014,Liu2016}.
Figure~\ref{fig:phase}(a) shows the variation of $|\Delta^{\eta}_{s}|$ with $\Delta_{z}$ for all four pairs of $(s,\eta)$ values and the corresponding Chern numbers have been listed in Tab. ~\ref{tab:Chern}.\\
\indent Here it is important to further introduce the spin Chern number $\mathcal{C}_{s}$ and the valley Chern number $\mathcal{C}_{v}$ as: 
\begin{eqnarray} 
\mathcal{C}_{s} = \frac{1}{2}(\mathcal{C}^{K}_{\uparrow}+\mathcal{C}^{K’}_{\uparrow}-\mathcal{C}^{K}_{\downarrow}-\mathcal{C}^{K’}_{\downarrow}) \label{spinCeq} \\ 
\mathcal{C}_{v} = \mathcal{C}^{K}_{\uparrow}-\mathcal{C}^{K’}_{\uparrow}+\mathcal{C}^{K}_{\downarrow}-\mathcal{C}^{K’}_{\downarrow}. \label{valleyCeq}
\end{eqnarray}
Based on $\mathcal{C}_{s}$ and $\mathcal{C}_{v}$, we can now label the various topological phases, namely the QSH phase and the QVH phases by varying $\Delta_{z}$. For $-\lambda_{SO} < \Delta_{z}<\lambda_{SO}$, the material exists in the QSH phase, characterized by $2\mathcal{C}_{s}=+2$ (given $\lambda_{SO}>0$) and $\mathcal{C}_{v}=0$. On the other hand, for  $\Delta_{z}< -\lambda_{SO}$ $(\Delta_{z}>\lambda_{SO})$, the material is in the QVH-I (QVH-II) phase having $2\mathcal{C}_{s}=0$ and $\mathcal{C}_{v}=+2 (-2)$.\\
\textbf{Topological Domain Walls.} 
We now move on to investigate the interface states at the topological domain walls between different phases~\cite{wang2014,top_dom_wall,zhang2020,Liu2016,SOC_DW_Abergel_2014,junzhu2016,yang2020}.
At the QVH-I/QSH interface, $\Delta \mathcal{C}^{\eta}_{s}$ is non-zero only for the  $(\uparrow,K’)$ and $(\downarrow,K)$ states and has opposite signs, thus yielding a pair of counter-propagating helical interface states. Hence at the domain wall between the QVH-I and QSH phases, we have spin-valley-momentum locked interface states, $(\uparrow,K’)$ and $(\downarrow,K)$ propagating in opposite directions. A similar analysis of the domain wall between the QVH-II and QSH phases yields counter-propagating $(\uparrow,K)$ and $(\downarrow,K’)$ interface states. There is a third possible scenario that arises when $\lambda_{SO}=0$, that is, no QSH phase exists and we will have a domain wall between the QVH-I and QVH-II phases. In this case $\Delta \mathcal{C}^{\eta}_{s}$ is non-zero for all the four possible states with opposite signs for the $K$ and $K’$ valley states.
Thus we expect four interface states with the $K$ valley states propagating in one direction and the $K’$ states in the opposite direction. One special feature of these interface states which makes them an attractive option for valley filtering is that the counter-propagating states are localised at the $K$ and $K’$ points and are immune to any back-scattering due to long range disorder~\cite{wang2014,cheng2016,junzhu2016}.\\
\textbf{Device structure.} 
The domain walls discussed above can be easily realized by creating a line junction between oppositely gated regions using a dual-split-gate structure shown in Figs.~\ref{fig:dual_gate}(a) and~\ref{fig:dual_gate}(b). The given configuration of voltage sources ensures that a voltage of $U\pm V$ is applied to the gates, where $V$ is responsible for modulating $E_{z}$ and $U$ allows us to control the channel electrochemical potential. Thus the channel has an additional on-site potential $U$ on every site realized via the addition of the Hamiltonian term $\hat{H}_{U} = \sum_{i\alpha}^{} U c_{i\alpha}^{\dagger}c_{i\alpha}
$. For the leads, we consider both $\lambda_{SO}$ and $\Delta_{z}$ to be zero. \\
\indent Given the width of the nanoribbon is $W_{o}$, we consider its y-coordinates to be in the range $\left[-\frac{W_{o}}{2},\frac{W_{o}}{2}\right]$. The dual-split-gates have a spacing of $W_{j}$, with the adjacent edges located at $y = \pm \frac{W_{j}}{2}$. We have modelled the spatial dependence of $E_{z}$ using the following analytical expression:
\begin{equation}\label{Ezeq}
E_{z}(y) = \frac{E_{zo}}{2} \left( {\bf{tanh}} \left( \frac{y-\frac{W_{j}}{2}}{\frac{W_{f}}{2}} \right) + {\bf{tanh}} \left( \frac{y+\frac{W_{j}}{2}}{\frac{W_{f}}{2}} \right)   \right),
\end{equation}         
where $E_{zo}$ is the maximum magnitude of the perpendicular electric field applied and $W_{f}$ is the fringing width of the electric field at the gate edges. In an actual setup $W_{f}$ has a direct dependence on the perpendicular spacing between the gates, which we assume to be $3.125nm$. Figures ~\ref{fig:phase}(b) and ~\ref{fig:phase}(c) illustrate the potential variation ($\mu_{i}\Delta_{z}$) along sublattices $A$ and $B$ for the two cases: $W_{j}=6.25nm$ (small $W_{j}$) and $W_{j}=20.8nm$ (large $W_{j}$) respectively. The above two cases differ in the fact that in the former case, the width of the region having $\Delta_{z}=0$ is negligible, whereas in the latter this width is large enough. Interestingly, exploiting the above potential variations, it is possible to have different cases of domain walls as depicted in Figs.~\ref{fig:dual_gate}(c) and~\ref{fig:dual_gate}(d), a deeper discussion of which will follow in the upcoming section. \\
%For the small $W_{j}$ case (Fig.~\ref{fig:domain_tis}(a)), for both the cases of without($\lambda_{SO}=0eV$) and with SO coupling($\lambda_{SO}=0.1eV$),  we expect a similar scenario i.e. QVH-I/QVH-II domain wall and hence four interface states within the line junction as shown in Fig.~\ref{fig:dual_gate}(c) because the junction is not wide enough to host a possible QSH region in presence of SO coupling. 
%Whereas when $W_{j}$ is large (Fig.~\ref{fig:domain_tis}(b)), in the presence of SO coupling, we have the expected QVH-I/QSH/QVH-II scenario with spatially separated spin-valley locked helical interface states (as shown in Fig.~\ref{fig:dual_gate}(d)) i.e. now counter-propagating states with same spin have negligible overlap between them and hence reduced chances of backscattering. 
%When SO coupling is absent, the junction region with zero $\lambda_{SO}$ and $\Delta_{z}$ has to be gapless and we get a QVH-I/semimetal/QVH-II scenario with all the four interface states now stretched out over this wider junction (Fig.~\ref{fig:domain_tis}(b)).
\begin{figure*}
    \centering
    \includegraphics[scale=0.55]{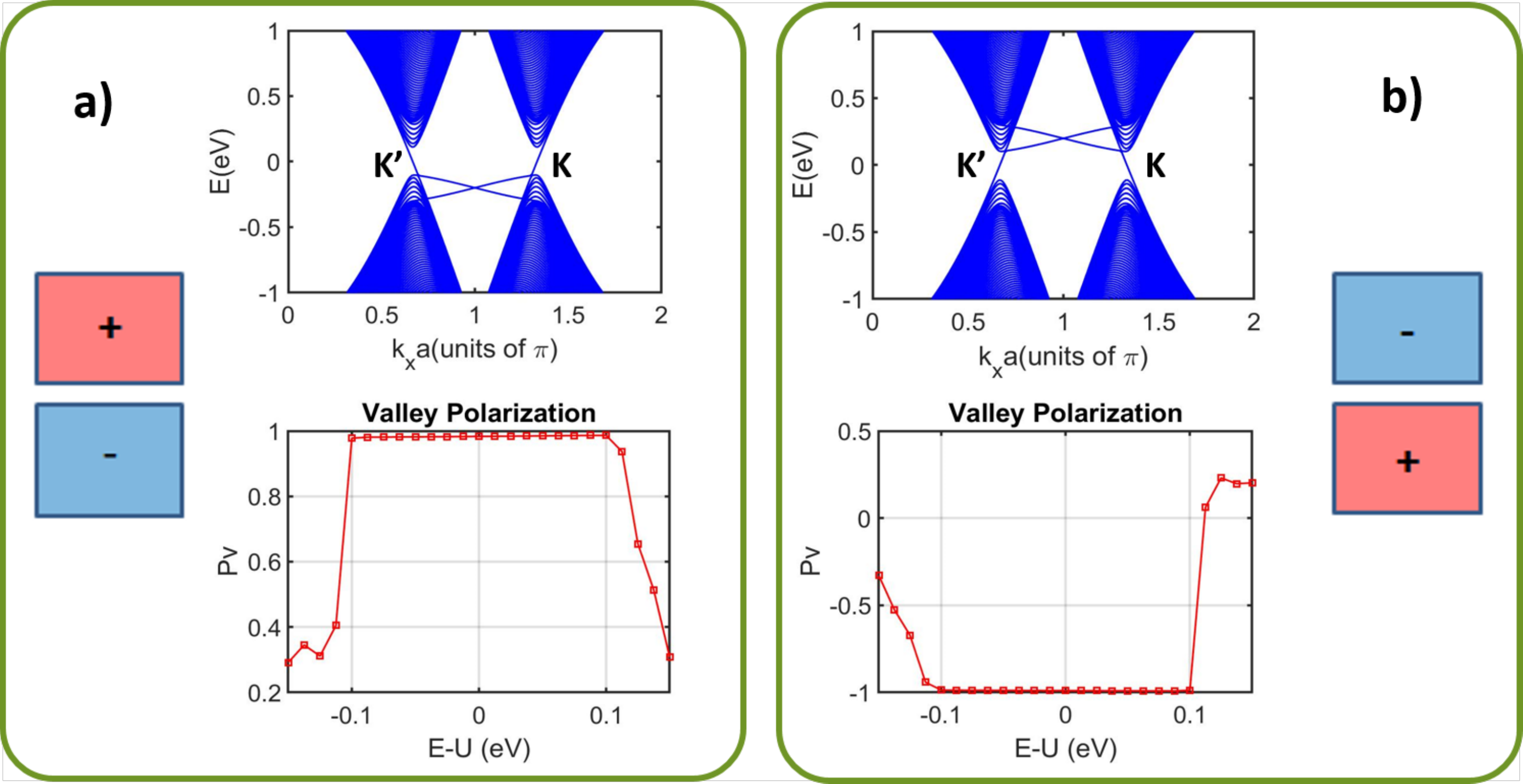}
    \caption{Gate voltage configuration, corresponding band structure and valley polarization of our proposed valley filter: $W_{j} = 6.25nm$ and $\lambda_{SO} = 0.1eV$ \textbf{a)} "$+-$" configuration gives a $K$-valley polarized filter for transmission from the left ($L$) to the right ($R$) lead, given the channel electrochemical potential lies within the band gap, where as suggested by the band structure, all the $K$-valley electrons are right-going and all $K'$-valley electrons are left-going. \textbf{b)} The  "$-+$" configuration makes it a $K'$-valley polarized filter as the interface states reverse their direction of propagation and now the $K$-valley electrons are left-going whereas the $K'$-valley electrons are right-going.
    }
    \label{fig:valley_filter}
\end{figure*}
\indent In all our calculations, we consider a zig-zag nanoribbon which is $180$ atoms wide, corresponding to a width $W_{o}=62.5nm$. The channel region $C$ as well as the leads $L$ and $R$ have the same width $W_{o}$, with the length $L_{o}$ of the channel region being 40nm i.e., $100$ atoms long, which is long enough such that the out-going carriers are completely valley polarized in the clean limit. The electrochemical potential of the leads ($E$) is fixed at $E = t/3$, corresponding to precisely seventy eight propagating modes, and that of the channel ($E-U$) is controlled by varying $U$. For all the numerical results that follow, unless explicitly mentioned, the width of the nanoribbon, length of the channel region $C$ and the electrochemical potential of the leads are fixed.\\
\indent Depicted in Fig.~\ref{fig:valley_filter} is the operation of our valley filter in the absence of any form of disorder, i.e., the clean limit.
The electrochemical potential of both the leads is fixed at $E$.
For this case: $W_{j}=6.25nm$, $E=t/3$, $\lambda_{SO}=0.1eV$ and $\Delta_{zo} = lE_{zo} = 0.2eV$(max. value of $\Delta_{z}$), thus creating a gap of $2\left(\Delta_{zo}-\lambda_{SO}\right)=0.2eV$.
The gap defined above corresponds to the gap of the QVH regions, given that the QSH region gap is independent of $\Delta_{zo}$ and depends only on $\lambda_{SO}$.
Both Fig.~\ref{fig:valley_filter}(a) and Fig.~\ref{fig:valley_filter}(b) corresponding to opposite gate configurations, show close to perfect valley polarization in the gap range $\left[ -0.1eV, 0.1eV \right]$.
Since the valley polarization plotted is for transmission from $L$ to $R$, whenever the $K$-valley electrons are right going, our filter is $K$-valley polarized as in Fig.~\ref{fig:valley_filter}(a) and similarly $K’$-valley polarization in Fig.~\ref{fig:valley_filter}(b). The fact that the valley polarization of the filter can be reversed by just reversing the signs of the gate voltages, makes this valley filter design quite convenient for practical purposes.
\subsection{Results}
\label{subsection:Results}
%Having described the setup and the model hamiltonian along with the relevant parameters involved, we next investigate, through our numerical calculations, the effect of intrinsic SOC ($\lambda_{SO}$) and the adjacent gate spacing ($W_{j}$) on the performance of our proposed valley filter subject to disorder.
%We begin by exploring the effect of varying $\lambda_{SO}$ for two different cases of $W_{j}$: 6.25nm (small $W_{j}$) and 20.8nm (large $W_{j}$).   
%Three different values of $\lambda_{SO}$ typically found in various 2D Xene materials have been considered: 0 meV(Graphene/Silicene), 40meV(Germanene) and 100meV(Stanene). Thus we consider six different combinations of $\lambda_{SO}$ and $W_{j}$.
\begin{figure*}
    \centering
    \includegraphics[scale=0.6]{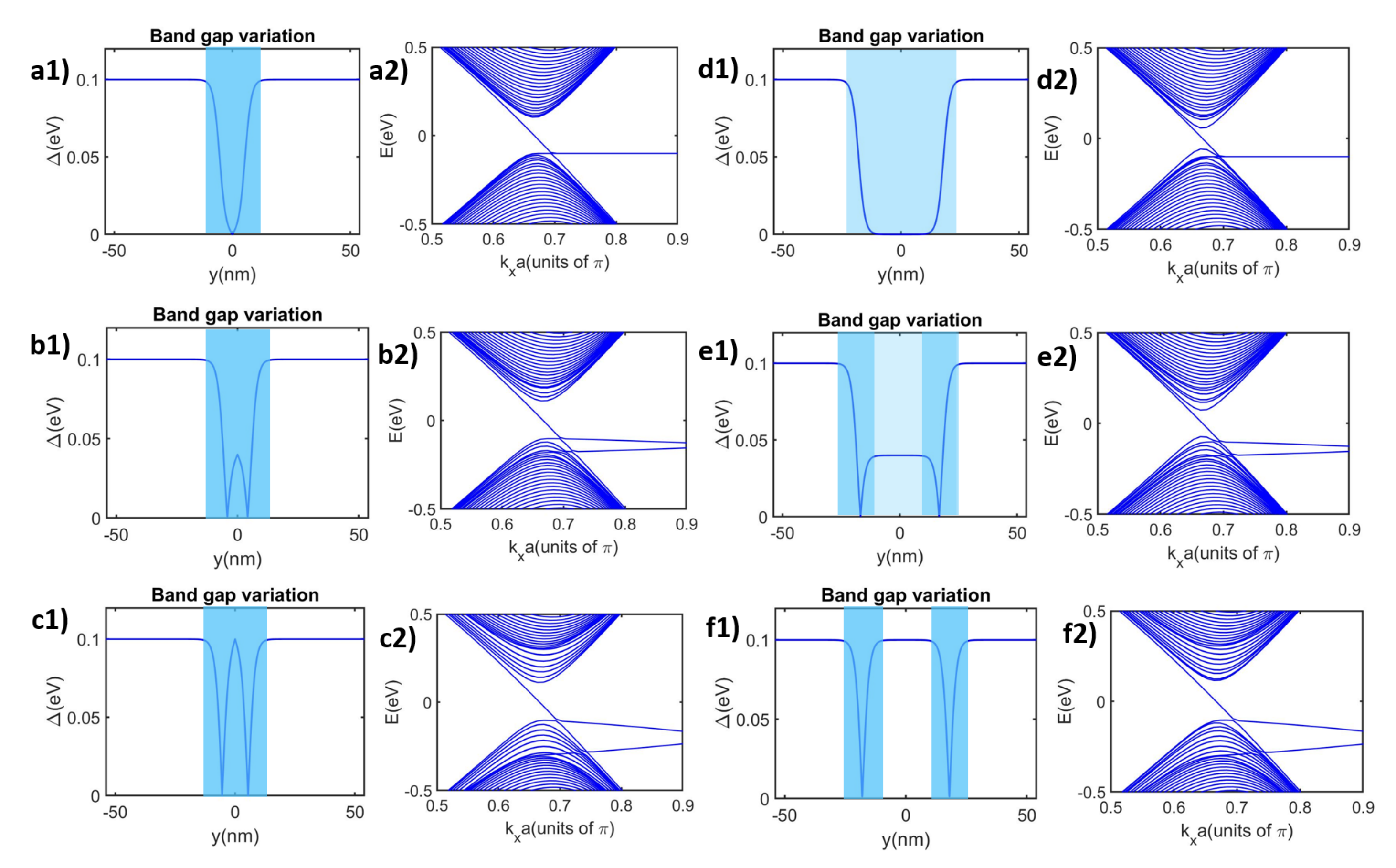}
    \caption{Spatial variation of local band gap and electronic band structure for different cases of $\lambda_{SO}$ and $W_{j}$ considered.
    \textbf{a1, a2)}  $W_{j} = 6.25nm$ and $\lambda_{SO}=0eV$
    \textbf{b1, b2)}  $W_{j} = 6.25nm$ and $\lambda_{SO}=0.04eV$
    \textbf{c1, c2)}  $W_{j} = 6.25nm$ and $\lambda_{SO}=0.1eV$
    \textbf{d1, d2)}  $W_{j} = 20.8nm$ and $\lambda_{SO}=0eV$
    \textbf{e1, e2)}  $W_{j} = 20.8nm$ and $\lambda_{SO}=0.04eV$
    \textbf{f1, f2)}  $W_{j} = 20.8nm$ and $\lambda_{SO}=0.1eV$
    }
    \label{fig:gap_band}
\end{figure*}
\begin{figure*}
    \centering
    \includegraphics[scale=0.6]{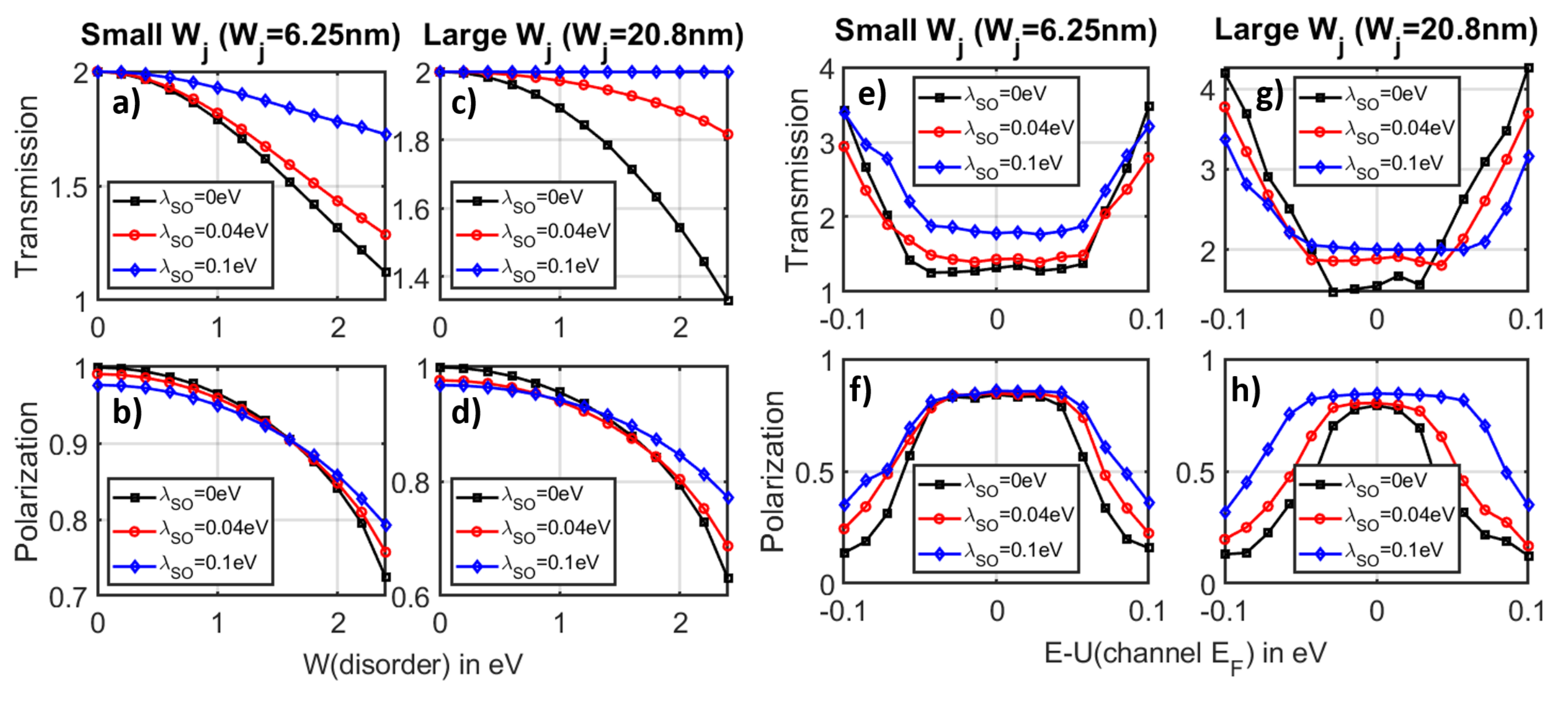}
    \caption{ Numerical results for different combinations of $W_{j}=6.25nm$, $20.8nm$ and $\lambda_{SO}=0eV$, $0.04eV$, $0.1eV$. In \textbf{(a)-(d)}, the results are plotted versus varying disorder strength $W$ for channel electrochemical potential, $E-U=0$. In \textbf{(e)-(h)}, the results are plotted versus varying $E-U$ over the band gap for $W =2eV$. The case having $W_{j}=20.8nm$ and $\lambda_{SO}=0.1eV$ clearly outperforms the rest.}
    \label{fig:main_results}
\end{figure*}

To ensure a fair comparison across all the cases, we assume the band gap of the bulk states in the QVH regions to be $0.2eV$, i.e., a gap range of $\left[ -0.1eV, 0.1eV \right]$, demanding $\Delta_{zo} = 0.1+\lambda_{SO}$, which implies that as $\lambda_{SO}$ is varied, the applied gate voltage $V$ also needs to be varied accordingly. Figure~\ref{fig:gap_band}(a1-f1) shows the spatial variation of local band gap along the nanoribbon width for six different combinations of $\lambda_{SO}$ and $W_{j}$ along with the corresponding band structures in Figure~\ref{fig:gap_band}(a2-f2). As indicated by the blue bands in  Fig.~\ref{fig:gap_band}(a1-f1), localized interface states appear at the locations where the band gap closes. The total transmission $T$ and valley polarization $Pv$ are plotted in Fig.~\ref{fig:main_results}, in two different ways: a) Disorder strength $W$ varied, for $E-U=0$, with $E=t/3$, as shown in Fig.~\ref{fig:main_results}(a-d) b) $E-U$ varied over the gap range $\left[ -0.1eV, 0.1eV \right]$ for $W=2eV$, as in Fig.~\ref{fig:main_results}(e-h). In the clean limit, within the gap range, we expect $Pv$ to be close to unity, and $T=2$, corresponding to two interface states, one for each spin.\\
\indent However for disorder strength $W=2eV$, $Pv$ deteriorates significantly, suggesting an increased inter-valley scattering between the helical interface states and hence a suppressed transmission due to increased backscattering. This does hold true for $E-U$ around zero, but when $E-U$ approaches the band edge, an enhancement in $T$ is observed indicating the bulk state assistance in inter-valley scattering, similar to what has been reported in~\cite{cheng2016}.
Thus the inter-valley mixing that is detrimental to the efficiency of our valley filter can happen either directly via backscattering between the helical interface states or indirectly assisted by the bulk states. The former is dominant when the electrochemical potential lies in the middle of the band gap and the latter takes over as the electrochemical potential approaches the band edge. \\
\indent Although there is no clear way to get rid of the bulk-assisted inter-valley scattering other than having a larger band gap, the direct one can be evaded by introducing a spatial separation between interface states of the same spin. The results in Figs.~\ref{fig:main_results}(c) and ~\ref{fig:main_results}(g) indeed confirm this. Unlike the other cases, as suggested by  Fig.~\ref{fig:main_results}(c), $T$ remains unchanged when $\lambda_{SO}=0.1eV$ and $W_{j} = 20.8nm$, even in the presence of very strong disorder ($W=2.4eV$) and does not degrade with $W$ as seen in other cases. Figure~\ref{fig:main_results}(g) shows that $T$ remains perfectly quantized when the electrochemical potential lies around the mid-gap before getting affected by the bulk states as we move closer to the band edge. Despite showing a mild degradation in $Pv$ with $W$ like the others, it still has the best valley polarization in the presence of strong disorder.\\
\indent The obtained results make more sense once the local band gap profiles in Fig.~\ref{fig:gap_band} are analyzed. Since the interface states appear at the locations where the gap closes, for the $W_{j}=6.25nm$ case, the four interface states co-exist and overlap with one another (Fig.~\ref{fig:gap_band}(a1-c1)). On the other hand when $W_{j}=20.8nm$, the interface states can now be hosted with spatial separation. However that alone is not enough to ensure spatial separation without any overlap. For example, in the case of $\lambda_{SO}=0eV$, the central line junction region is semi-metallic and gapless thus leading to the interface states spreading out uniformly over the entire central region instead of being localized as depicted in Fig.~\ref{fig:gap_band}(d1).\\
\indent A wide semi-metallic region also leads to bulk states, thus decreasing the available band gap as suggested by Fig.~\ref{fig:gap_band}(d2). This has a detrimental effect on the valley filter performance, with the device now not only having a diminished operational energy range, but also a degrading valley polarization within this range, as confirmed by Fig.~\ref{fig:main_results}(h), thus highlighting the role of bulk-assisted inter-valley scattering in deteriorating the filtering efficiency.
Similarly when $\lambda_{SO}=0.04eV$ there exists a low band gap QSH region in the line junction. Although the interface states are hosted at different locations as depicted in Fig.~\ref{fig:gap_band}(e1), they can still interact with one another via tunneling through the low band gap central region, thus leading to inevitable back-scattering. Even in this case, some bulk states enter the band gap as shown in Fig.~\ref{fig:gap_band}(e2) and further deteriorate the valley filter performance.\\
\indent However, when $\lambda_{SO}=0.1eV$, the gap is free from bulk states as the QSH region has a gap equal to that of the QVH regions and any possible tunneling between the interface states is suppressed owing to the larger bandgap central region. Thus the back-scattering is strongly suppressed even in presence of strong disorder and we get a perfectly quantized transmission $T=2$ around the mid-gap. However the bulk-assisted inter-valley scattering still persists and hence the degradation of $Pv$ with $W$. \\ 
\begin{figure*}
    \centering
    \includegraphics[scale=0.5]{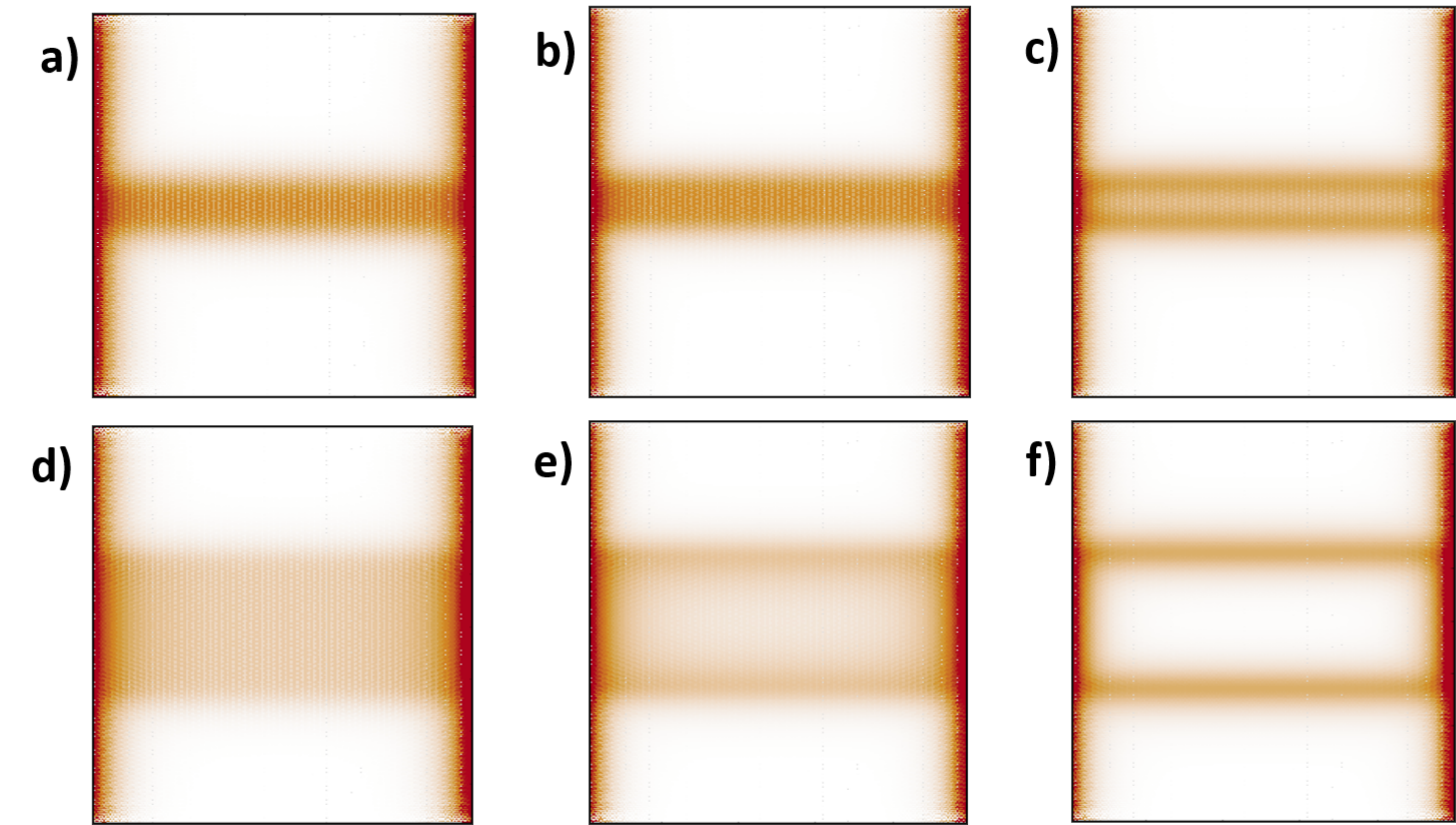}
    \caption{LDOS profile for different cases of $\lambda_{SO}$ and $W_{j}$ considered. The channel electrochemical potential $E-U=0$ and disorder strength $W=0$
    \textbf{a)}  $W_{j} = 6.25nm$ and $\lambda_{SO}=0eV$
    \textbf{b)}  $W_{j} = 6.25nm$ and $\lambda_{SO}=0.04eV$
    \textbf{c)}  $W_{j} = 6.25nm$ and $\lambda_{SO}=0.1meV$
    \textbf{d)}  $W_{j} = 20.8nm$ and $\lambda_{SO}=0eV$
    \textbf{e)}  $W_{j} = 20.8nm$ and $\lambda_{SO}=0.04eV$
    \textbf{f)}  $W_{j} = 20.8nm$ and $\lambda_{SO}=0.1eV$
    }
    \label{fig:ldos}
\end{figure*}
\indent To further validate the arguments made above, the LDOS in the absence of disorder, for $E-U=0$, has been plotted for all the cases in Fig.~\ref{fig:ldos}. The spatial variation of the interface states is illustrated for varied $\lambda_{SO}$, when $W_{j}$ is small, in Fig.~\ref{fig:ldos}(a-c) and similarly in Fig.~\ref{fig:ldos}(d-f) for large $W_{j}$. The LDOS plots confirm our previous claims of the interface states being spatially separated only for the case with $\lambda_{SO}=0.1eV$ and $W_{j}=20.8nm$ (Fig.~\ref{fig:ldos}(f)). Thus our simulation results shed light on the possibility of achieving an improved valley filter performance, with a perfectly quantized $T$, by spatially separating the interface states with the same spin through a suitable chosen gate configuration.  We now look into the important aspect of optimizing the proposed valley filter device.

\subsection{Device Optimization}
\begin{figure*}
    \centering
    \includegraphics[scale=0.6]{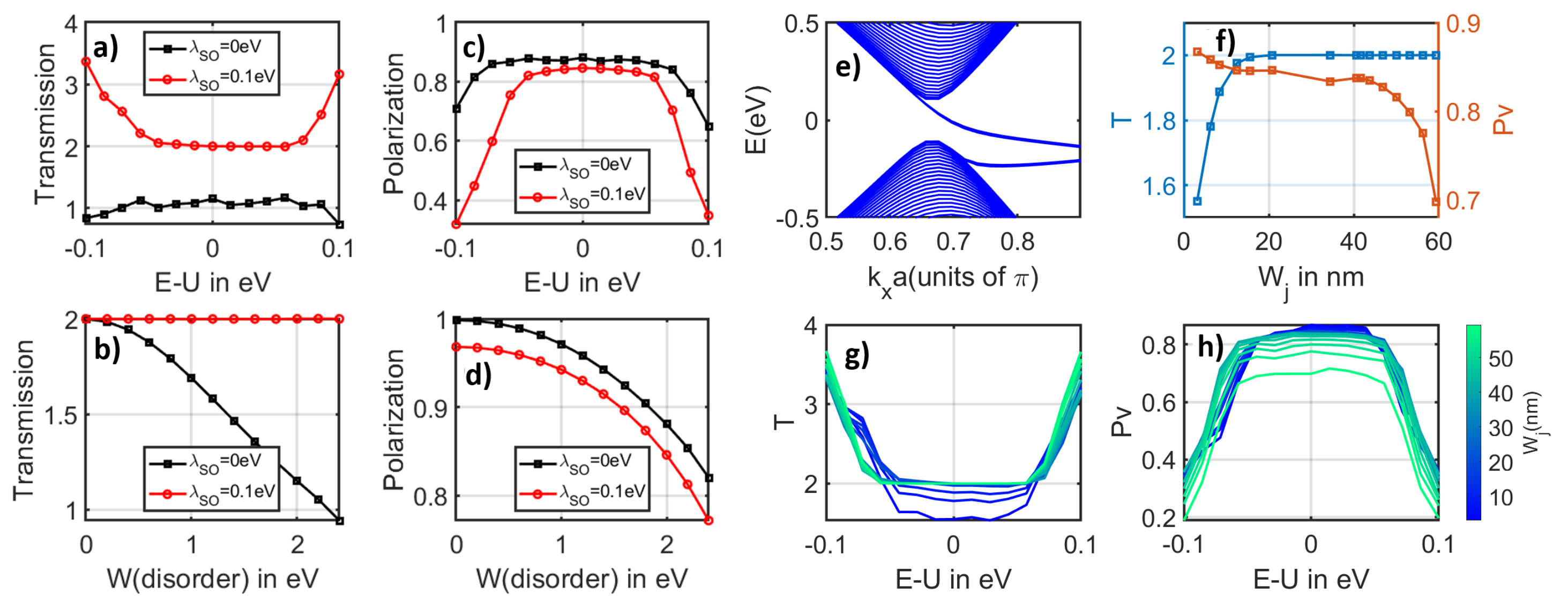}
    \caption{Comparison of valley filtering efficiency for 
    the cases with ($\lambda_{SO}=0.1eV$) and without ($\lambda_{SO}=0eV$) SO coupling when subjected to the same dual gate voltage.
    \textbf{a,c)} $T$ and $Pv$ versus $E-U$ for $W=2eV$
    \textbf{b,d)} $T$ and $Pv$ versus $W$ for $E-U=0eV$ 
    \textbf{e)} Band structure for $W_{j}\approx 60nm$, the interface states move towards the $X$-point of the Brillouin zone
    \textbf{f)} Variation of $T$ and $Pv$ w.r.t $W_{j}$ for $W=2eV$ and $E-U=0eV$. $T$ shows droop at low $W_{j}$ because of increased tunneling between the interface states whereas $Pv$ shows droop at high $W_{j}$ as the interface states move away from the $K$ and $K'$ points.
    \textbf{g,h)} Variation of $T$ and $Pv$ w.r.t $E-U$ for different values of $W_{j}$ as depicted by the color bar and $W=2eV$. For intermediate values of $W_{j}$ we expect the best performance in terms of both $T$ and $Pv$.
    }
    \label{fig:yso_effect}
\end{figure*}

As mentioned before, a good electrical valley filter should have a large valley polarization along with a large enough total current in order to achieve a large valley current. In other words,  perfect valley polarization with a very small amount of current is of no practical utility to any of the valleytronic applications. Thus exploiting the topological robustness and dissipation-less nature of spatially separated interface states with the introduction of intrinsic SO coupling is the key to achieve improved valley filtering.\\
\indent Our valley filter can be optimized by considering the following two crucial  requirements: Firstly, the QSH region should be well-gapped and wide enough to ensure that the spatially separated interface states do not overlap with each other. This is necessary to ensure that backscattering is negligible and $T$ is perfectly quantized even in the presence of strong non-magnetic disorder. Secondly, the QVH regions also need to be well-gapped and wide enough to ensure that the interface states do not spread out to the nanoribbon edges. Ensuring this is critical to maintain the valley-polarized character of the interface states so that the valley filter produces significant valley polarization. \\
% \subsection{Optimal $\lambda_{SO}$}
\textbf{Optimal $\bf{\lambda_{SO}}$.}
\indent So far, the comparison between the different cases was made after considering the same bulk band gap of the QVH region and this requires different perpendicular fields $E_{zo}$ for different values of $\lambda_{SO}$.  In Fig.~\ref{fig:main_results}, we notice that when $\lambda_{SO}=0eV$, the valley filter performs the best when the adjacent gates are as close as possible, whereas for $\lambda_{SO}=0.1eV$ the valley filter efficiency is enhanced when the adjacent gates are far apart to ensure spatial separation of interface states. Hence for $\lambda_{SO}=0eV$ we consider $W_{j}=6.25nm$ whereas when $\lambda_{SO}=0.1eV$ we increase $W_{j}$ to 20.8nm to have the best possible valley filter. Similar to Fig.~\ref{fig:gap_band}, we consider $\Delta=0.1eV$ for $\lambda_{SO}=0.1eV$ case and this requires $E_{zo} = 5V/nm$ given $l=0.4${\AA}, which is practically achievable. For the same $E_{z}$ and $l$, in the $\lambda_{SO}=0eV$ case we can achieve $\Delta=0.2eV$.\\ 
\indent The transmission plots in Fig.~\ref{fig:yso_effect}(a,b) clearly show the superiority of $\lambda_{SO}=0.1eV$ case in its robustness to back-scattering. 
 However, as suggested by Fig.~\ref{fig:yso_effect}(c,d), $Pv$ degrades a bit in this case compared to the $\lambda_{SO}=0eV$ case because of a smaller band gap leading to increased bulk-assisted inter-valley scattering. Thus when the same gate voltage is applied, addition of SO coupling to host spatially separated interface states helps us achieve a perfectly quantized $T$ but at the cost of a reduced $Pv$. 
\indent In the above comparison, we considered $E_{zo}=5V/nm$ where for the $\lambda_{SO}=0.1eV$ case we had $\Delta=0.1eV$ and hence we achieved reasonably good valley filtering whereas for a lower $E_{zo}$ say, $3V/nm$, the performance degrades due to a very small $\Delta$ of $0.02eV$. The bare minimum one needs to ensure is that $\lambda_{SO}<E_{zo}/l$, in order to assure the existence of the QSH/QVH domain walls, and hence the interface states.
%Thus for a given $E_{zo}$ one would prefer a smaller $\lambda_{SO}$ to have a larger band gap for the QVH regions so that the interface states are perfectly valley polarized and  are less affected by the bulk states.
%However that would mean a small band gap QSH region and as suggested by results for $\lambda_{SO}=0.04eV$ in Fig.~\ref{fig:main_results}, the interface states are not robust to back-scattering due to the possibility of tunneling between them and the bulk states corresponding to the QSH region enter the band gap thus enhancing bulk-assisted inter-valley scattering. 
Hence the best case would be when the QSH as well as the QVH regions have the same gap implying $\lambda_{SO} = lE_{zo}/2$ for a given $E_{zo}$. Thus for $E_{zo}=5V/nm$ and $l=0.4${\AA}, $\lambda_{SO} = lE_{zo}/2 = 0.1eV$ is the optimal one to get the best yields as valley filter given the interface states are well separated after choosing a suitable $W_{j}$.\\
% \subsection{Optimal $W_{j}$}
\textbf{Optimal $\bf{W_{j}}$.}
One might expect that a larger $W_{j}$ should ensure a greater degree of robustness for the interface states. However, since we have considered a fixed width for the nanoribbon ($W_{o}=62.5nm$), invariably increasing $W_{j}$ implies that the widths of the QVH regions keep decreasing and the interface states move closer to the nanoribbon edges. As a consequence, the interface states start resembling the conventional QSH edge states as shown in Fig.~\ref{fig:yso_effect}(e), in which $W_{j}\approx 60nm$, which then results in diminishing $Pv$ as they move towards the $X$-point of the Brillouin zone. On the other hand, decreasing $W_{j}$ also brings the interface states closer from which they interact with each other via tunneling, thus further degrading $T$ via back-scattering. Thus $W_{j}$ needs to be chosen such that the interface states are neither too close to each other nor too far apart that they come in the vicinity of the nanoribbon edges.\\
\indent To quantify the above arguments, we consider the spatial extent of the interface states on either side of the domain wall given by the damping length, $\zeta = \hbar v_{f}/2\Delta = \sqrt(3)at/4\Delta$~\cite{wang2014,yongxu2019}. One thing to note is that this expression is valid for an abrupt topological phase transition. However, in our case, it is rather smooth over a fringing width of $W_{f}=3.125nm$. Hence the actual $\zeta$ would be slightly larger than what we estimate. It is clear that $\zeta$ is inversely proportional to $\Delta$, thus maintaining our earlier conclusion that $\Delta$ needs to be large enough for all the QSH as well as QVH regions, so that the interface states are well confined. For the electric field we consider ($E_{zo} = 5V/nm$) the corresponding optimal $\lambda_{SO}$ was $0.1eV$, thus $\Delta=0.1eV$ for the QSH as well as QVH regions. The $\zeta$ calculated for our case comes out to be $2.25nm$ which is in fact smaller than $W_{f}=3.125nm$ and hence we choose $\zeta=3.125nm$. Considering exponential variation of the wavefunction along the $y$- direction on either sides of the domain wall, 96\% of the wavefunction remains confined within a distance i.e., three times the damping length $\zeta$, for our case $3\zeta\approx 9.5nm$.\\
\indent Based on the above calculations, we can conclude for the spatially separated interface states to be decoupled from each other, we require $W_{j}> 2\times 9.5=19nm$ and to ensure that they do not spread out all the way into the nanoribbon edges $W_{j} < W_{o}-2\times 9.5=43.5nm$ is required. Hence choosing $19nm < W_{j} < 43.5nm$, ensures the desirable valley filtering performance both in terms of $T$ and $Pv$. \\
\indent Our analytical calculations are also validated by the simulation results in Fig.~\ref{fig:yso_effect}(f), where $T$ and $Pv$ are plotted against varying $W_{j}$ for $E-U=0$ and $W=2eV$.  The total transmission $T$ remains constant at $2$, as we lower $W_{j}$ before declining rapidly as $W_{j}$ goes below $20nm$.  
On the other hand, the valley polarization $Pv$ decreases steadily as we increase $W_{j}$, before showing a sudden droop once $W_{j}$ goes past $40nm$. Thus our choice of $W_{j}=20.8nm$ when $E_{zo}=5V/nm$ and $\lambda_{SO}=0.1eV$, as suggested by both our analytical calculations and simulation results, is optimal in terms of both $T$ and $Pv$. 
The variation of $T$ and $Pv$ over the entire band gap, for different values of $W_{j}$, is illustrated in Fig.~\ref{fig:yso_effect}(g,h). 
One important thing to note is that in our above optimization strategy, we were constrained by the maximum electric field $E_{zo}$ we can apply and the nanoribbon width $W_{o}$, unconstrained by which, we can further enhance the valley filter performance. \\

\section{Conclusion}
\label{section:Conclusion}
In conclusion, we proposed a design for an all-electrical robust valley filter that utilizes topological interface states hosted on monolayer group-IV 2D-Xene materials with large intrinsic spin-orbit coupling.
In contrast with conventional QSH edge states localized around the $X$-points, the interface states appearing at the domain wall between topologically distinct phases are either from the $K$ or $K’$ points, making them suitable prospects for serving as valley-polarized channels. 
We showed that the presence of a large band-gap quantum spin Hall effect facilitates the spatial separation of the spin-valley locked helical interface states with the valley states being protected by spin conservation, leading to robustness against short-range non-magnetic disorder. By adopting the scattering matrix formalism on a suitably designed device structure, valley-resolved transport in the presence of non-magnetic short-range disorder for different 2D-Xene materials was analyzed in detail.
Our numerical simulations confirm the role of spin-orbit coupling in achieving an improved valley filter performance with a perfect quantum of conductance attributed to the topologically protected interface states. Our analysis further elaborated clearly the right choice of material, device geometry and other factors that need to be considered while designing an optimized valleytronic filter device.
% We finally discussed the experimental feasibility of our proposal. 
We believe that our work opens the door for researching the utility of the 1D topological conducting channels hosted in the monolayer 2D-Xene bulk for possible applications in valleytronics and spintronics.

\section{Methods}
\label{section:Metodology}
We use the software package \texttt{"KWANT"}~\cite{kwant} for calculating the transfer matrix $\tau$, which gives us the transmission amplitude $\tau_{k_{2},k_{1}}$ from the $k_{1}$ state of lead $L$ to the $k_{2}$ state of lead $R$.  
The inter-valley($T_{KK’}$ and $T_{K’K}$) and intra-valley($T_{KK}$ and $T_{K’K’}$), transmission coefficients can be calculated using the following formula~\cite{honeycomb_modes,beenakker2007}:
\begin{equation}\label{Tkkeq}
T_{K_{2}K_{1}} = \sum_{k_{1} \in K_{1}}^{} \sum_{k_{2} \in K_{2}}^{} |\tau_{k_{2},k_{1}}|^{2}
\end{equation}
Now the transmission coefficient corresponding to scattering into the $K$-valley in lead $R$ is given by $T_{K} = T_{KK}+T_{KK’}$ and similarly $T_{K’} = T_{K’K}+T_{K’K’}$.
Once we have the valley-resolved transmission coefficients($T_{K}$ and $T_{K’}$), we can now calculate the valley polarization $Pv$ and the total transmission $T$~\cite{honeycomb_modes,beenakker2007}, the two important metrics to evaluate the performance of a valley filter.
\begin{equation}\label{Pveq}
Pv = \frac{T_{K}-T_{K’}}{T_{K}+T_{K’}}
\end{equation}
\begin{equation}\label{Teq}
T = T_{K}+T_{K’}.
\end{equation}
To evaluate the performance of our valley filter in actual experimental conditions, we subject it to short-range Anderson non-magnetic disorder which does not break TRS and this can be done by introducing random on-site potential for each site. 
Despite the Anderson disorder not reflecting all the
potential valley-mixing mechanisms in experimental samples, it does provide a computationally efficient means to model the inter-valley scattering and allows us to examine the effect of varying parameters such as $\lambda_{SO}$ and $W_{j}$ on the valley filter performance.
This is achieved by adding the term  $\hat{H}_{W} = \sum_{i\alpha}^{} \epsilon_{i} c_{i\alpha}^{\dagger}c_{i\alpha}$ to the channel Hamitonian, with $\epsilon_{i}$ being randomly distributed in the interval $\left[-\frac{W}{2},\frac{W}{2}\right]$ where $W$ is the disorder strength~\cite{honeycomb_modes,cheng2016,junzhu2016}.
For each value of $W$, fifty different random disorder configurations are considered and the results are averaged over all the configurations.\\
% \indent In all our calculations, we consider a zig-zag nanoribbon which is $180$ atoms wide, corresponding to a width $W_{o}=62.5nm$. The channel region $C$ as well as the leads $L$ and $R$ have the same width $W_{o}$, with the length $L_{o}$ of the channel region being 40nm i.e., $100$ atoms long, which is long enough such that the out going carriers are completely valley polarized in the clean limit. The electrochemical potential of the leads ($E$) is fixed at $E = t/3$ and that of the channel ($E-U$) is controlled by varying $U$. For all the numerical results above, unless explicitly mentioned, the width of the nanoribbon, length of the channel region $C$ and the electrochemical potential of the leads are fixed.\\
\indent We must remark at this stage that only static impurities inside the channel are considered within the above approach. An unexplored frontier in terms of understanding the stability of topological edge/interface states is the inclusion of momentum relaxing dephasing \cite{7571106,PhysRevApplied.8.064014,doi:10.1063/1.5023159,Praveen,Duse_2021} phase breaking processes \cite{DANIELEWICZ1984239,PhysRevB.75.081301,PhysRevB.98.125417} and ultimately inelastic scattering processes \cite{Aniket_Singha,Aniket_Singha_2}, that can be well facilitated by using the Keldysh non-equilibrium Green's function approach \cite{Datta}. 

% \begin{acknowledgment}
%  The research and development work undertaken in the project under the Visvesvaraya Ph.D Scheme of the Ministry of Electronics and Information Technology (MEITY), Government of India, is implemented by Digital India Corporation (formerly Media Lab Asia). This work is also supported by the Science and Engineering Research Board (SERB), Government of India, Grant No. Grant No. STR/2019/000030, the Ministry of Human Resource Development (MHRD), Government of India, Grant No. STARS/APR2019/NS/226/FS under the STARS scheme. 
% \end{acknowledgment}

\section*{Data Availability}
The data generated from this work is available from the authors upon request. 
\begin{acknowledgements}
The research and development work undertaken in the project under the Visvesvaraya Ph.D Scheme of the Ministry of Electronics and Information Technology (MEITY), Government of India, is implemented by Digital India Corporation (formerly Media Lab Asia). This work is also supported by the Science and Engineering Research Board (SERB), Government of India, Grant No. Grant No. STR/2019/000030, the Ministry of Human Resource Development (MHRD), Government of India, Grant No. STARS/APR2019/NS/226/FS under the STARS scheme.
% Please use ``The authors thank \ldots'' rather than ``The
% authors would like to thank \ldots''.

% The author thanks Mats Dahlgren for version one of \textsf{achemso},
% and Donald Arseneau for the code taken from \textsf{cite} to move
% citations after punctuation. Many users have provided feedback on the
% class, which is reflected in all of the different demonstrations
% shown in this document.

\end{acknowledgements}

\section*{Author Contributions}
BM and KJ conceived the idea. KJ performed all numerical simulations. All authors contributed in analyzing the results and writing the paper.

\section*{Competing Interests}
The authors declare that there are no competing interests.

% \begin{suppinfo}

% % This will usually read something like: ``Experimental procedures and
% % characterization data for all new compounds. The class will
% % automatically add a sentence pointing to the information on-line:

% \end{suppinfo}

\bibliography{achemso-demo}

\end{document}